\documentclass[jcp,aip,10pt,floatfix,twocolumn,showpacs]{revtex4-1} 
\usepackage{color,soul} 
\usepackage{graphicx}
\usepackage{amsmath, amsthm, amssymb}
\usepackage{amsthm}
\usepackage{multirow} 
\usepackage[normalem]{ulem}
\newcommand{\be}{\begin{equation}}
\newcommand{\ee}{\end{equation}}
\newcommand{\bea}{\begin{eqnarray}}
\newcommand{\eea}{\end{eqnarray}}
\begin{document} 
\title{Aggregation of flexible polyelectrolytes: Phase diagram and dynamics.} 
\author{Anvy Moly Tom} 
\email{anvym@imsc.res.in} 
\affiliation{The Institute of Mathematical Sciences, C.I.T. Campus, 
Taramani, Chennai 600113, India} 
\affiliation{Homi Bhabha National Institute, Training School Complex, Anushakti Nagar, Mumbai 400094, India}
\author{R. Rajesh} 
\email{rrajesh@imsc.res.in}  
\affiliation{The Institute of Mathematical Sciences, C.I.T. Campus, 
Taramani, Chennai 600113, India} 
\affiliation{Homi Bhabha National Institute, Training School Complex, Anushakti Nagar, Mumbai 400094, India}
\author{Satyavani Vemparala} 
\email{vani@imsc.res.in} 
\affiliation{The Institute of Mathematical Sciences, C.I.T. Campus, 
Taramani, Chennai 600113, India} 
\affiliation{Homi Bhabha National Institute, Training School Complex, Anushakti Nagar, Mumbai 400094, India}
\date{\today}
\begin{abstract}
Similarly-charged polymers in solution, known as polyelectrolytes, are known to form aggregated structures in the presence of oppositely charged counterions. Understanding the dependence of the equilibrium phases and the dynamics of the process of aggregation on parameters such as backbone flexibility and charge density of such polymers is crucial for insights into various biological processes which involve biological polyelectrolytes such as protein, DNA etc., Here, we use large-scale coarse-grained molecular dynamics simulations to obtain the phase diagram of the aggregated structures of flexible charged polymers and characterize the morphology of the aggregates as well as the aggregation dynamics, in the presence of trivalent counterions. Three different phases are observed depending on the charge density: no aggregation, a finite bundle phase where multiple small aggregates coexist with a large aggregate, and a fully phase separated phase. We show that the flexibility of the polymer backbone causes strong entanglement between charged polymers leading to additional time scales in the aggregation process.  Such slowing down of the aggregation dynamics results in the exponent, characterizing the power law decay of the number of aggregates with time, to be dependent on the charge density of the polymers. These results are contrary to those obtained for rigid polyelectrolytes, emphasizing the role of backbone flexibility.
\end{abstract}

\maketitle

\section{Introduction} 
Polyelectrolytes (PEs) are macromolecules with ionizable groups which release 
counterions when the PEs are
dissolved in solutions,  rendering the backbone of the polymer chain charged. Examples of PEs include biological 
polymers such as DNA, RNA, actin, virus etc.~\cite{PhysRevE.71.061928,BIP:BIP360311305,BLOOMFIELD1996334},  as well as synthetic polymers such as sulphonated polystyrene, polyacrylic acid etc.~\cite{mortimer1991synthetic}.  PEs have a wide range of applications such as gene therapy~\cite{wolfert1996characterization,blocher2016complex,guo2010enhanced,mansouri2006characterization}, drug coating~\cite{kawashima1985novel,yang2010mesoporous,schmidt2010electrically,berg2006controlled}, water purification~\cite{bolto2007organic,shannon2008science,kawamura1991effectiveness,tripathi2013functional}, color removal~\cite{hong2006removal,aravind2010treatment,ramesh1994hydrophobic,bidhendi2007evaluation}, 
paper making~\cite{mortimer1991synthetic,wagberg2002engineering} etc.
The dynamical and structural properties of the PEs, critical for their applications, are crucially dependent on the
conformational phases that the PEs may assume depending on a variety of conditions and parameters of the system. 
These phases are primarily determined by the competition between the repulsive electrostatic
interactions among the like-charged monomers of the PE chains and the entropy of the free counterions. Depending on
the relative dominance, 
free counterions
may condense onto the polymer backbone~\cite{manning1969,manning1972,kamcev2016}, 
renormalizing the charge density, and facilitate effective short-ranged attractive interactions between
monomers. In the dilute limit of a single PE chain in isolation, the effective attractive interactions can result in 
extended, bead-necklace, and collapsed conformations depending on the charge density of PE chain and temperature of the system.~\cite{Kremer1995,Kremer1993,winkler98,brilliantov98,Melnikov1999,limbach03,Dobrynin2005,Chertovich2016,Chertovich2016a,jayasree2009equilibrium,limbach2003single,chang2003strongly,dobrynin1996cascade,lyulin1999computer,budkov2017,budkov2015,anoop11}

At finite densities of PE chains,  the effective attractive interactions among the PE chains can lead to aggregation, in addition to individual collapsed phases. Understanding counterion mediated aggregation 
of charged polymers is very relevant as the aggregation of biopolymers such as DNA and actin has been implicated to play an important role in biological functions such as cell scaffolding, DNA packaging, 
cytoskeletal organization~\cite{Bloomfield, Needleman, Claessens, Huber-actin, Sarah-actin,Angelini03,Janmey, Wong-Rev}. In addition to biological polymers, recent studies have show that aggregation of synthetic polymers is crucial in their ability to function as biomimetic and functional materials~\cite{Kouwer, Bertrand,Fletcher,Stricker}. The primary questions of interest regarding aggregation of PE chains are: Does aggregation always lead to complete phase separation or sometimes result in finite bundles? What is the
morphology of the aggregate phases? Does the aggregation dynamics depend on the details of the system or is it universal? Are the effective interactions driving the aggregation of multiple PE chains similar to those responsible for the collapse of a single flexible PE chain? 

These questions have been addressed in experiments, in theoretical studies as well as in large-scale simulations involving rigid rod-like PE chains (RLPE chains). Several experiments~\cite{Bloomfield,tang,sedlak,tanahatoe, borsali,zribi,butler,bordi,Muhlrad,huang2014,tanahatoe1997}, computer simulations and theoretical analyses~\cite{ray,perico2,Zhou-DFT,ermoshkin,Broedersz,perico2,manning3,yethiraj97,yethiraj09,manning2014,Fazli1,Fazli2,savelyev,Holm2007,Holm2010,stevens1,stevens2,pietronave,jensen,luan,bruinsma,diehl2001} have shown that RLPE chains aggregate in the presence of multivalent counterions, though some controversy exists in the case of monovalent counterions~\cite{sedlak,tanahatoe,borsali,ermi,zhang,ray,huang2014,perico2,zribi,butler,arenzon1,arenzon2,solis,jensen,stevens1,stevens2,diehl2001,allahyarov,savelyev,anoop12,anvy16}. However, a clear resolution is lacking as to whether the thermodynamic equilibrium conformation of such self-organization of the PE chains is a complete phase separated state in which all PE chains aggregate~\cite{Ha1998,huang1994early,Borue1990},  or coexistence of multiple finite aggregates~\cite{bloomfield1991condensation,podgornik1994parametrization,bloomfield1996dna,tang1996polyelectrolyte,tang1997opposite,PinacusPRE2005}. Computer simulations of a system of 61 RLPE chains, with moderate values of charge density and relatively short simulation time scales, suggest that finite-sized bundles is the thermodynamic equilibrium state~\cite{Holm2007,Holm2010,Fazli2,Fazli1}. Large scale simulations also suggest that the final thermodynamic state could crucially depend on the charge density~\cite{Holm2007,Holm2010}, volume fraction or salt content~\cite{tamashiro2006rayleigh}. The formation of finite bundles has been attributed to  kinetic barriers arising because the aggregate cannot achieve charge neutralization due to the steric and short-ranged electrostatic interactions~\cite {PinacusPRE2005}. However, in our previous studies of RLPE chains~\cite{anoop12, anvy16}, we observed the formation of a phase separated state as the thermodynamically equilibrium state, over long simulation time scales, contrary to the previously mentioned studies. The aggregation dynamics, in the case of RLPE chains, was studied by looking at the scale free power law evolution of size of the aggregates (or decrease in number of aggregates, $N(t)\sim t^{-\theta}$).  From large-scale simulations and modeling the evolution of the aggregate size distribution through Smoluchowski coagulation equation, we obtained $\theta=2/3$, in contrast to the exponent value of $\theta=1$, obtained in other studies~\cite{Fazli2,Fazli1}. This power law exponent was shown to be independent of charge density of the polymers, valency of the counterions, density, and length of the PE chain, and the aggregation of RLPE chains was shown to be diffusion-limited~\cite{anvy16}. 

While the aggregation and self-organization of RLPE chains are reasonably well understood, much remains to be explored regarding 
the aggregation mechanism of flexible PE (FPE) chains, whose conformational flexibility can introduce additional time scales and barriers in the aggregation dynamics. Biological polymers such as proteins are essentially FPEs and understanding the role of aggregation of FPEs is relevant as  protein-protein disordered aggregates are implicated in many neurodegenerative diseases~\cite{fink1998protein,de2012intrinsic}.Viscoelastic and scattering measurements on a variety of FPE solutions, including both biopolymers like \textit{Aggrecan} and synthetic copolymers,  have revealed that beyond a certain concentration of the FPE chains, signature of entanglement is clearly observed via change in scaling of correlation lengths~\cite{di2004structure, colby2010structure,dobrynin2008theory,mckee2006solution,di2007dynamic,dou2006charge,chitanu2000static,ermi1998domain}. Theoretical studies have also shown that for flexible unconstrained chains, formation of macroscopic aggregates via phase separation is the lowest energy state~\cite{huang2002}.  

The flexibility of  FPE chains can introduce additional kinetic barriers due to the disentanglement and subsequent entanglement  required for incorporating new FPE chains into an existing aggregate, altering the aggregation dynamics as compared to RLPE chains. It is possible that during the process of disentanglement to include a new FPE chain,  the existing  aggregate may fragment into individual PE chains, further complicating the aggregation dynamics. Another aspect of importance is to understand the underlying effective attractive interactions that play a dominant role in aggregation of like-charged FPE chains and whether these interactions are similar to those in the  collapsed phase of a single flexible 
PE chain. In recent work~\cite{tom2016mechanism,tom2016regimes} we have shown that the collapsed regime of a single FPE chain (in either good or poor solvent conditions) is composed of multiple sub-regimes. These sub-regimes are characterized by different scaling exponents in the relation between radius of gyration and the effective Bjerrum length of the PE chain, especially when the PE chain is strongly charged. Among existing theories to explain the counterintuitive collapse of a charged PE chain~\cite{brilliantov98,pincus98,Kardar1999,delaCruz2000,Cherstvy2010,muthukumar2004,arti14}, we identified counterion fluctuation theory~\cite{brilliantov98} to be the correct theory and modified it suitably to account for the existence of several sub-regimes in the collapsed phase of a single PE chain. It would be interesting to explore whether similar sub-regimes exist for strongly charged PE chains in their aggregated regime and whether counterion fluctuation theory still holds good for aggregated structures. It can be envisaged that when the strongly charged PE chains self-assemble into an aggregated structure, the counterions are no longer bound to the specific FPE chain but move freely within the aggregate of multiple PE chains and this scenario may not be very different from a collapsed regime of a single long collapsed PE chain. 

In this paper, we study the  equilibrium conformations and dynamics of aggregates of highly charged FPE chains, using molecular dynamics (MD) simulations (model and MD details in Sec.~\ref{sec:model}). In Sec.~\ref{sec:phase}, we discuss the effects of backbone flexibility of FPE chains on the equilibrium phases in the presence of trivalent counterions and contrast them with the results for RLPE chains with rigid backbones. We also characterize the morphology of the aggregates in detail in Sec.~\ref{sec:morphology} and study the role of conformation of a single FPE chain in emergence of kinetic barriers that affect the aggregation dynamics. In Sec.~\ref{sec:dynamics}, we characterize the dynamics of aggregation in the phase separated phase by measuring the power law exponent describing the 
decreasing in the number of aggregates. Finally, in Sec.~\ref{sec:summary} we provide a detailed discussion of our results. 

\section{Methods \label{sec:model}} 

We consider a system of $N=100$ PE chains, each one  consisting of 
$N_m=30$ monomers of charge $+e$, and corresponding number of  neutralizing counterions of charge 
$-Ze$, where $Z$ is the valency of the counterion. In the present study, we only consider trivalent counterions 
($Z=3$).  The PE chains are modeled using a bead spring model~\cite{anvy16,anoop12},
where the monomers are connected by harmonic springs with the interaction potential,
\begin{equation}
U_{bond}(r_{ij})=\frac{1}{2}k(r_{ij}-b)^2,
\end{equation}
where $k$ is the spring constant and $b$ is the equilibrium bond length. The flexibility of the chain is maintained by not including a three-body bond bending interaction.  

The non-bonded particles interact through $6$--$12$ Lennard Jones potential,
\begin{equation}
U_{LJ}(r_{ij})=4\epsilon \left[\left(\frac{\sigma}{r_{ij}}\right)^{12}-\left(\frac{\sigma}{r_{ij}}\right)^6\right],
\end{equation}
where $r_{ij}$ is the distance between particles $i$ and $j$,
$\epsilon$ is the minimum of the potential  and $\sigma$ 
is the inter-particle distance at which the potential is zero. The parameters $\epsilon$ and $\sigma$ are the same for all
pairs of particles. The Lennard Jones potential is smoothly cutoff at a distance $r_c$, 
which has been chosen to be 
$\sigma$, so that the interaction between all the particles is purely repulsive, implicitly mimicking good solvent conditions.

The electrostatic interaction among all pairs of particles is given by
Coulomb interaction,
\begin{equation}
U_{c}(r_{ij})=\frac{q_iq_j}{4\pi\epsilon_0 r_{ij}},
\label{eq.1}
\end{equation}
where $q_i$ and $q_j$ are the charges of the $i^{\mathrm{th}}$ and the
$j^{\mathrm{th}}$ particles, which can take values $e$ or $-Ze$, and $\epsilon_0$ is the permittivity. 

The strength of the charge density along the PE chain is parameterized by a 
dimensionless quantity $A$:
\begin{equation}
A=\frac{\ell_{B}}{b},
\label{eq.4}
\end{equation}
where $\ell_{B}$ is the Bjerrum length, the length scale below which 
electrostatic interactions dominate thermal energy~\cite{Russel}, and is defined as
\begin{equation}
\ell_{B}=\frac{e^{2}}{4\pi\epsilon_0 k_{B}T},
\label{eq.5}
\end{equation}
where $k_{B}$ is the Boltzmann constant and $T$ is temperature. 
Larger the value of $A$, stronger the electrostatic interactions. In simulations,
$A$ may be changed either by changing the permittivity $\epsilon_0$ or changing the
value of the charge $e$.

The PE chains and the corresponding counterions  are placed in a 
box of linear size $L$ with periodic boundary conditions.  The equations of motion are integrated in time using
the molecular dynamics (MD) simulation package LAMMPS~\cite{lammps1,lammps2} using a time step of 0.001 at temperature $T=1$ maintained
through a  Nos\'{e}-Hoover thermostat~\cite{nose,hoover}. 
Details of the interaction parameters are given in Table~\ref{table1}. 
The long-ranged electrostatic interactions are calculated using Particle-Particle/Particle-Mesh (PPPM) technique~\cite{Hockney}.
\begin{table}
\caption{Parameters used in the simulations. }
\label{table1}
\begin{center}
\begin{tabular}{cc}
\hline
\hline
Parameters&value\\ \hline
$\sigma$&1\\
$b$&1.12$\sigma$\\
$\epsilon$&1\\
$r_c$&$\sigma$\\
$T$&1\\
$k$&500\\
\hline
\end{tabular}
\end{center}
\end{table}

We perform two kinds of simulations which differ from each other in their initial conditions. The first set of NVT simulations
are used to obtain the different phases that the aggregates may exist in and the complete
number density ($\rho$)-$A$ phase diagram. The initial conditions for these sets of simulations were obtained by performing a NPT ($P=1$, $T=1$) simulation of a system of randomly dispersed PE chains with $A=3.57$ until all the 100 PE chains aggregate into
a single aggregate. This single aggregate is then evolved in time for different values of $A$ and number density $\rho$ for $10^7$ steps to
ascertain the stability and morphology of the resultant aggregates. These simulations will be referred to as ``reverse simulations".
In the second set of NVT simulations, the initial condition is one where all the PE chains and the counterions are dispersed uniformly thoughout the simulation box. To ensure that the initial conformations of the PE chains are well dispersed throughout the simulation box, we set the charge density of the PE chains to a very small value ($A=0.22$), much less than that of a critical charge density above which counterions condense onto the PE chains ($A_c \approx 1$) and equilibrate the system. Twenty random configurations, which are temporally well separated are chosen for further simulations in which the value of charge density of the FPE chains, $A$, is varied.  Here we keep the number density fixed 
($L=129$ and $\rho=1.4 \times 10^{-3}$) and vary $A$.  For each value of $A$ considered, $20$ independent simulations are performed, each for $10^7$ steps. 
These simulations will be referred to as ``forward simulations".  

\section{Results  \label{sec:results}}

\subsection{Equilibrium phases and phase diagram \label{sec:phase}}

We first identify the equilibrium phases and  phase diagram of a system of flexible PE chains, as a function of PE chain charge density $A$ and number density $\rho$. This could be done using forward simulations, starting with a dispersed set of PE chains and observe the aggregated structures that evolve with time. However, it would be difficult to conclusively distinguish between finite bundle formation and fully phase separated phases due to very long equilibration time scales. Setting an arbitrarily fixed equilibration time~\cite{Holm2007,Holm2010} can potentially lead to erroneous conclusions regarding the phase diagram. To avoid this problem and construct the phase diagram, we study the stability of a fully phase separated aggregate structure by performing reverse simulations at various values of $A$ and $\rho$.

\begin{figure}
\includegraphics [width=\columnwidth]{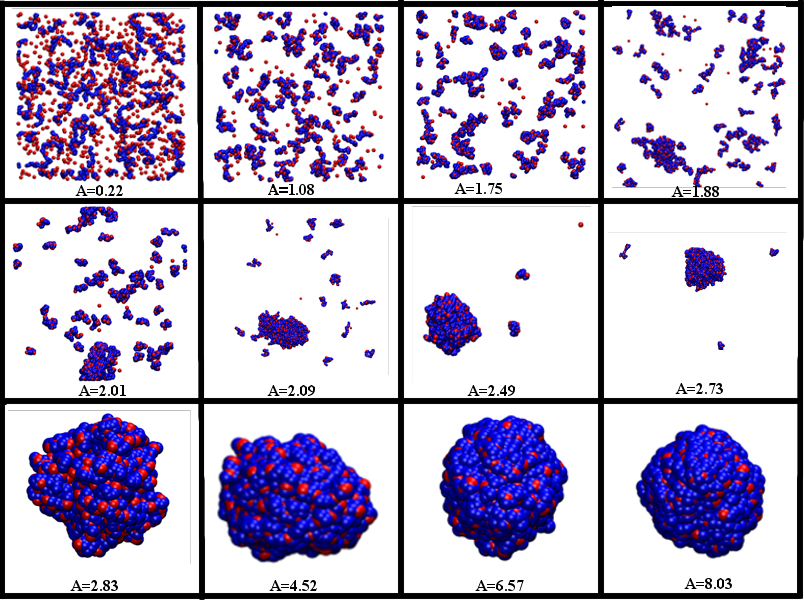}
\caption {Reverse simulations: snapshots of the system at the end of $10^7$ MD steps for different values of charge density $A$ and  at fixed 
number density $\rho=1.43\times10^{-3}$. The monomers and   counterions are colored in blue and red respectively. The snapshots are not to scale. }
\label{Fig:Aggphases}
\end{figure}

The snapshots of the system  at the end of $10^7$ steps are shown in Fig.~\ref{Fig:Aggphases} for different values of $A$
and fixed number density $\rho=1.43\times10^{-3}$. For a given value of $\rho$, we find that the system may exist in three different phases. 
These phases correspond to the aggregate being (a) completely fragmented (no aggregation), (b) partially fragmented (finite bundles characterized by the presence of a single large aggregate and 
multiple small aggregates), or 
(c) intact (phase separation).  

\begin{figure}
\includegraphics [width=\columnwidth]{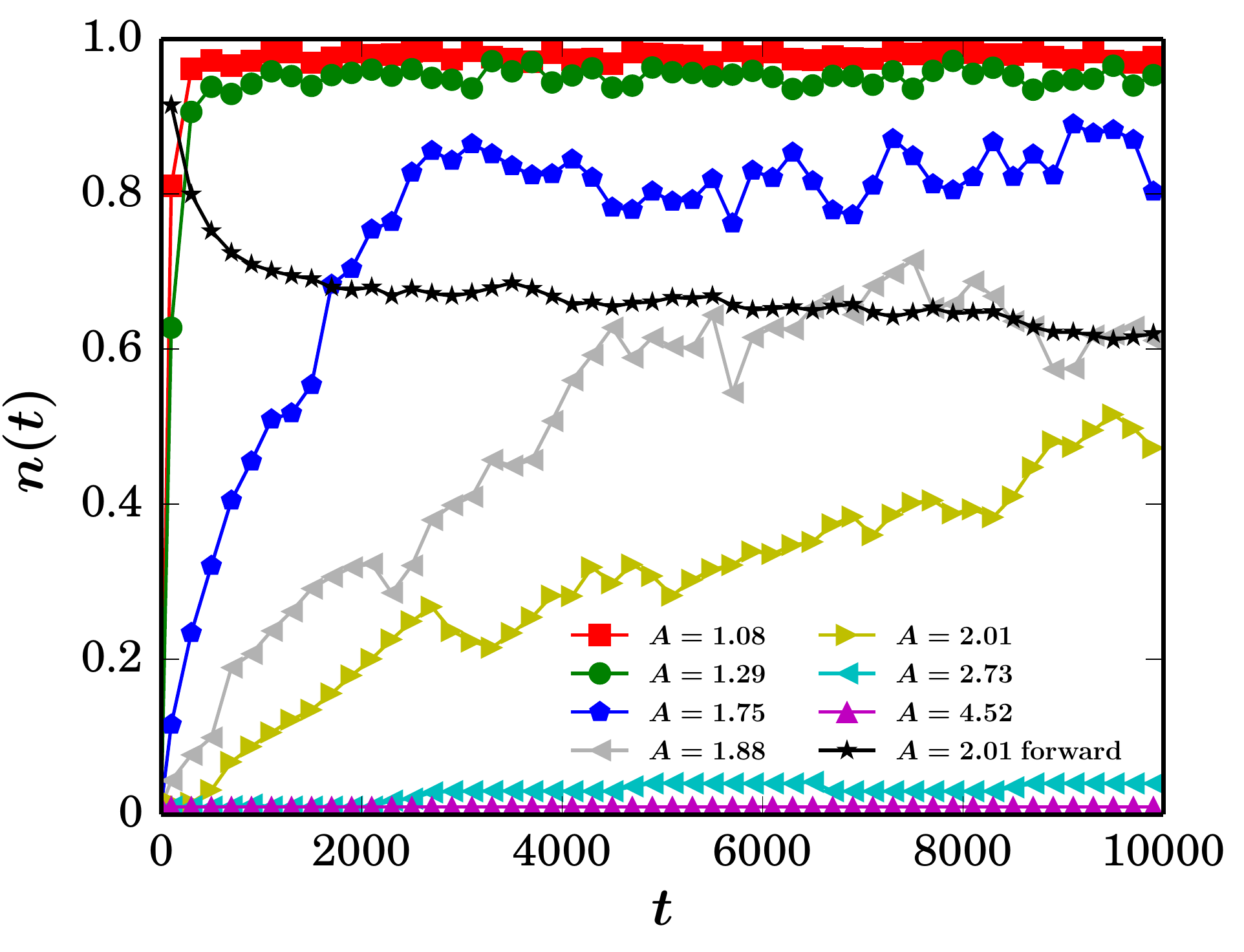}
\caption { The variation of the fraction of aggregates, $n(t)$ with time for different values of the charge density $A$ and fixed  
number density $\rho=1.43\times10^{-3}$ in reverse simulations. The initial condition is one where all PE chains are in one aggregate corresponding to $n(0)=N^{-1}=0.01$, where
$N$ is the total number of PE chains. For intermediate values of $A$, $n(t)$ attains a time-independent constant value 
between $.01$ and $1$, corresponding to the existence of finite-sized bundles. To show equilibration, $n(t)$ for $A=2.01$ in forward simulations, averaged over twenty initial conditions, is also shown. }
\label{Fig_nt}
\end{figure}

To show the stability of the three phases, the equilibration of the system as well as independence of the final state
on the initial conditions have to be checked. We do so by monitoring the fraction of aggregates $n(t)$, defined as the ratio of number of aggregates at any time $t$ to total number of PE chains in the system ($N=100$).  Two PE chains are said to form an aggregate of size two if the
distance between any two monomers belonging two different PE chains is
less than $2 \sigma$. The same definition is generalized to an
aggregate of size $m$. The variation of $n(t)$ with time in shown in Fig.~\ref{Fig_nt}. From the figure, three different regions are observed, consistent with Fig.~\ref{Fig:Aggphases}. 
For $A<1.28$, the initial single aggregate of size 100 fragments completely and $n(t)$ increases from $0.01$ to nearly $1$.  For $1.29<A<2.73$, within the first half of the simulation time,  $n(t)$ reaches a steady state value between 0 and 1. From these results, we identify two critical values of charge density denoted by $A_1\approx 1.29$ and $A_2 \approx 2.73$ above which finite bundles  and phase separated phases appear respectively. We check that
in this range of $A$ (for $A=2.01$) values, the steady state values of $n(t)$ are independent of the initial conditions by doing a forward simulation in which
all the PE chains are uniformly distributed within the simulation box.  The steady state values from these runs (averaged over 20 initial conditions) as well as from the reverse simulations tend to the same value at large times, showing that the system is equilibrated. For values of $A$ larger than 2.73, the initial single aggregate remains intact throughout the simulation time scale suggesting that for these values of $A$ complete phase separation is the equilibrium phase. We also checked that the fully phase separated phase (for $A=8.03$) is stable by confirming that  two aggregates of
size 50 each merge at large times to form a single aggregate of size 100.  

\begin{figure}
\includegraphics [width=\columnwidth]{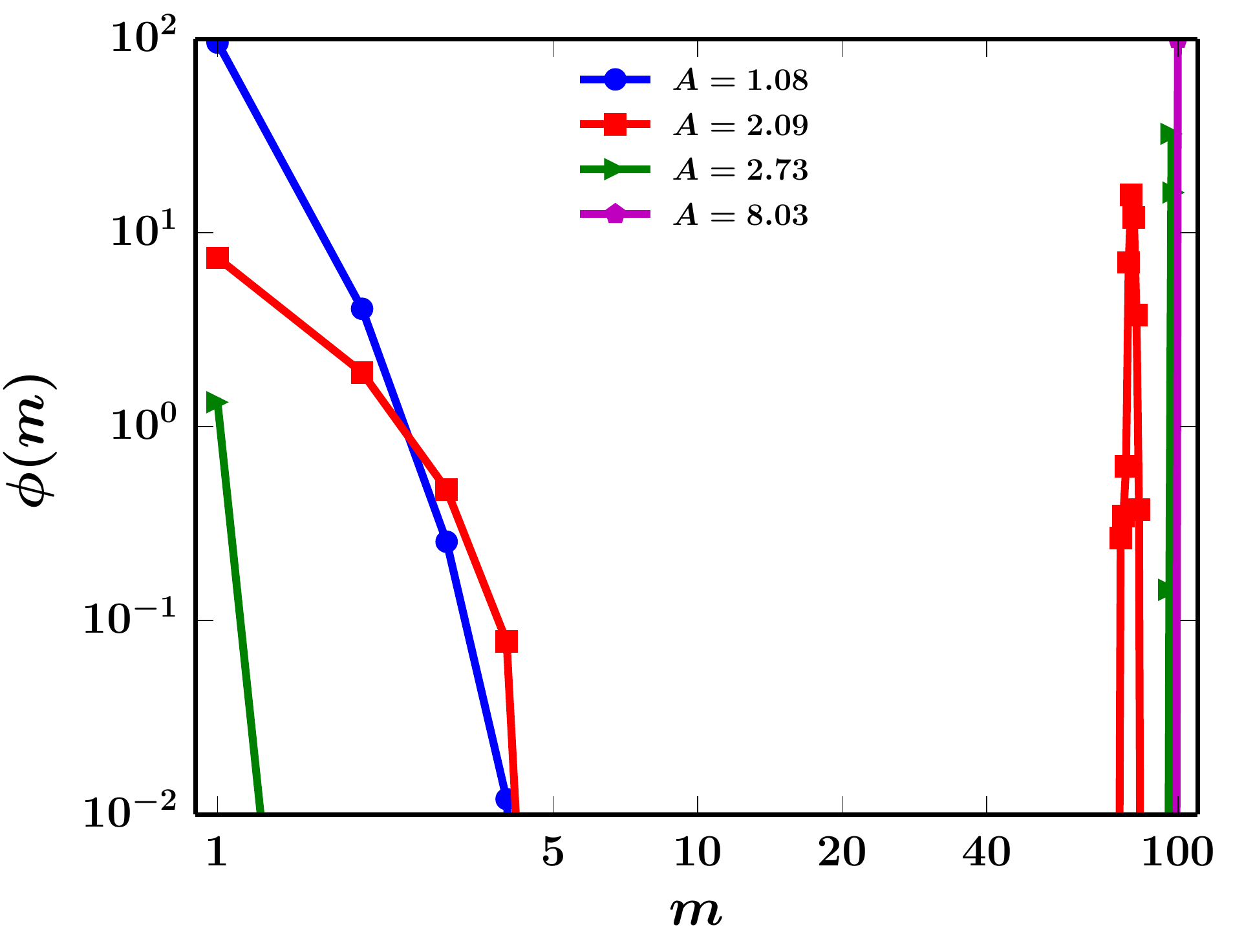}
\caption {$\phi(m)$, the fraction of PE chains in an aggregate of  size $m$ as a function of $m$ for different values of $A$. The data is for number density  $\rho=1.43\times10^{-3}$.}
\label{Fig:mpm}
\end{figure}

Characterizing the finite bundle phase only in terms of $n(t)$ is incomplete, as it does not distinguish between many small-size aggregates coexisting with a large aggregate,  or a system consisting of only aggregates of small size. To understand more about the aggregate size distribution, we define a parameter $\phi(m)= mN(m)$, where 
$N(m)$ is the density of aggregates of size $m$. $\phi(m)$ is the fraction of PE chains contained in aggregates of size $m$.
The variation of $\phi(m)$ with $m$ for different values of $A$ is shown  in Fig.~\ref{Fig:mpm}. The data is obtained by averaging
over the production run. For $A=1.08$, which is in the fully fragmented phase, it can be seen that the peak of the distribution is at 1, showing that the system is made of aggregates of very small size, almost all of them made of single PE chain. For two values of $A=2.09, 2.73$ in the finite bundle phase, the distribution is inhomogeneous and consists of two parts: a peak at large values of $m$ and a decaying distribution for small values of $m$.  This shows that the
finite bundle phase consists of the co-existence of a single large aggregate and multiple small-size aggregates. On the other hand, for the fully phase separated phase ($A=8.03$), the peak of the distribution is at 100, as expected for a stable aggregate of size $100$. The aggregate size distribution, along with the data for the total number of aggregates $n(t)$ shown in Fig.~\ref{Fig_nt}, clearly demonstrates the existence as well as the nature of the finite bundle phase, albeit for a range of $A$ values.

\begin{figure}
\includegraphics [width=\columnwidth]{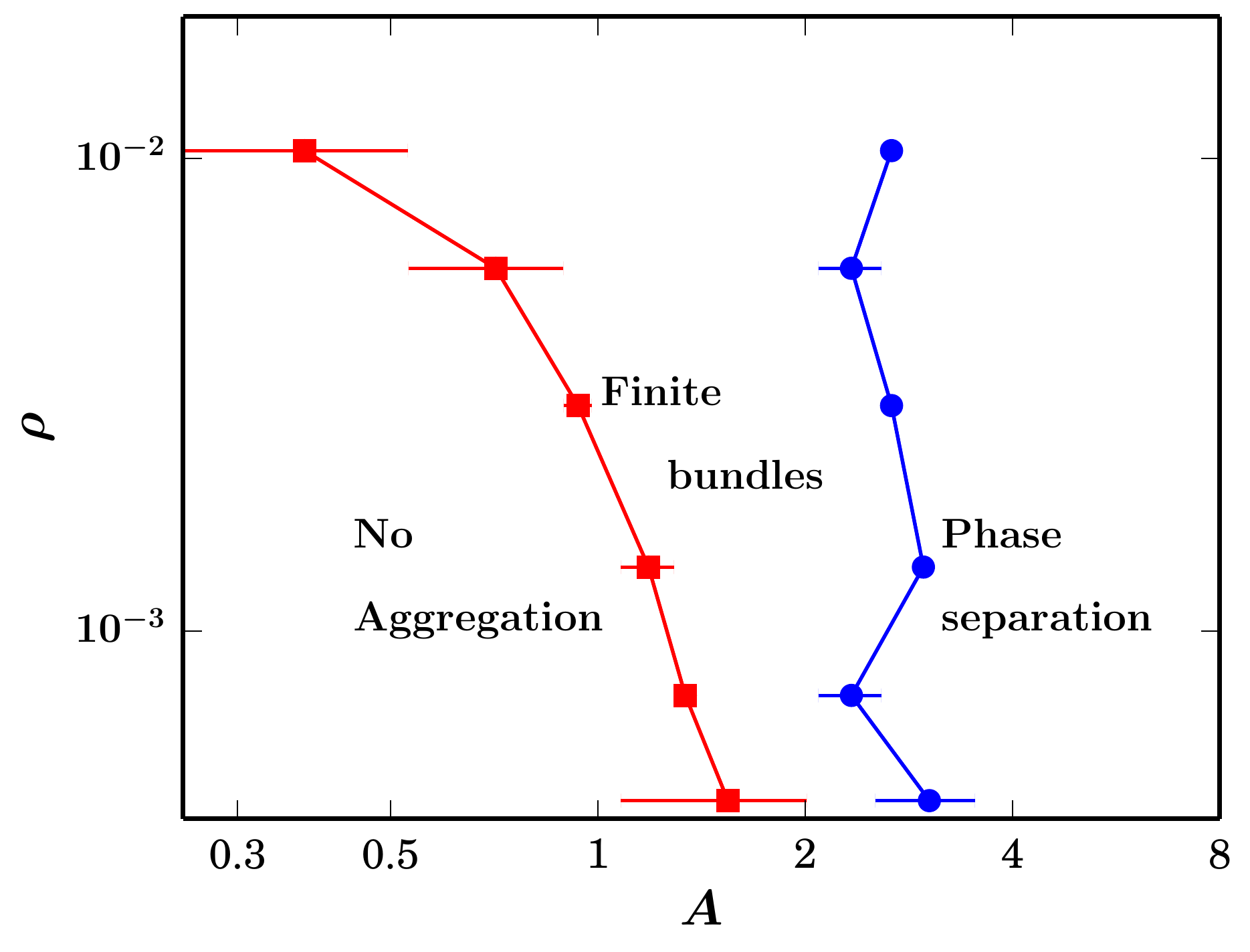}
\caption {Phase diagram in the $\rho$-$A$ plane. There are three phases: no aggregation, finite-sized bundles and completely phase separated phase. }
\label{Fig:phase_diagram}
\end{figure}

The critical values $A_1$ and $A_2$, which characterize the beginning and end of the finite bundle phase,
are expected to depend on the number density $\rho$. To construct the phase diagram in the $\rho$-$A$ plane, we performed extensive simulations for many values of $\rho$ and the results are shown in Fig.~\ref{Fig:phase_diagram}. The approximate transition points are identified by measuring $n(t)$ and  we label the region $0.95 <n(t)<.02$ as the finite bundle phase. From Fig.~\ref{Fig:phase_diagram}, within numerical error, the transition from the finite bundle phase to the fully phase separated phase appears to be independent of the number density for the range of densities considered in our simulations, and  hence
is presumably driven only by electrostatic interactions.

\subsection{Morphology of aggregates\label{sec:morphology}}

\begin{figure}
\includegraphics[width=\columnwidth]{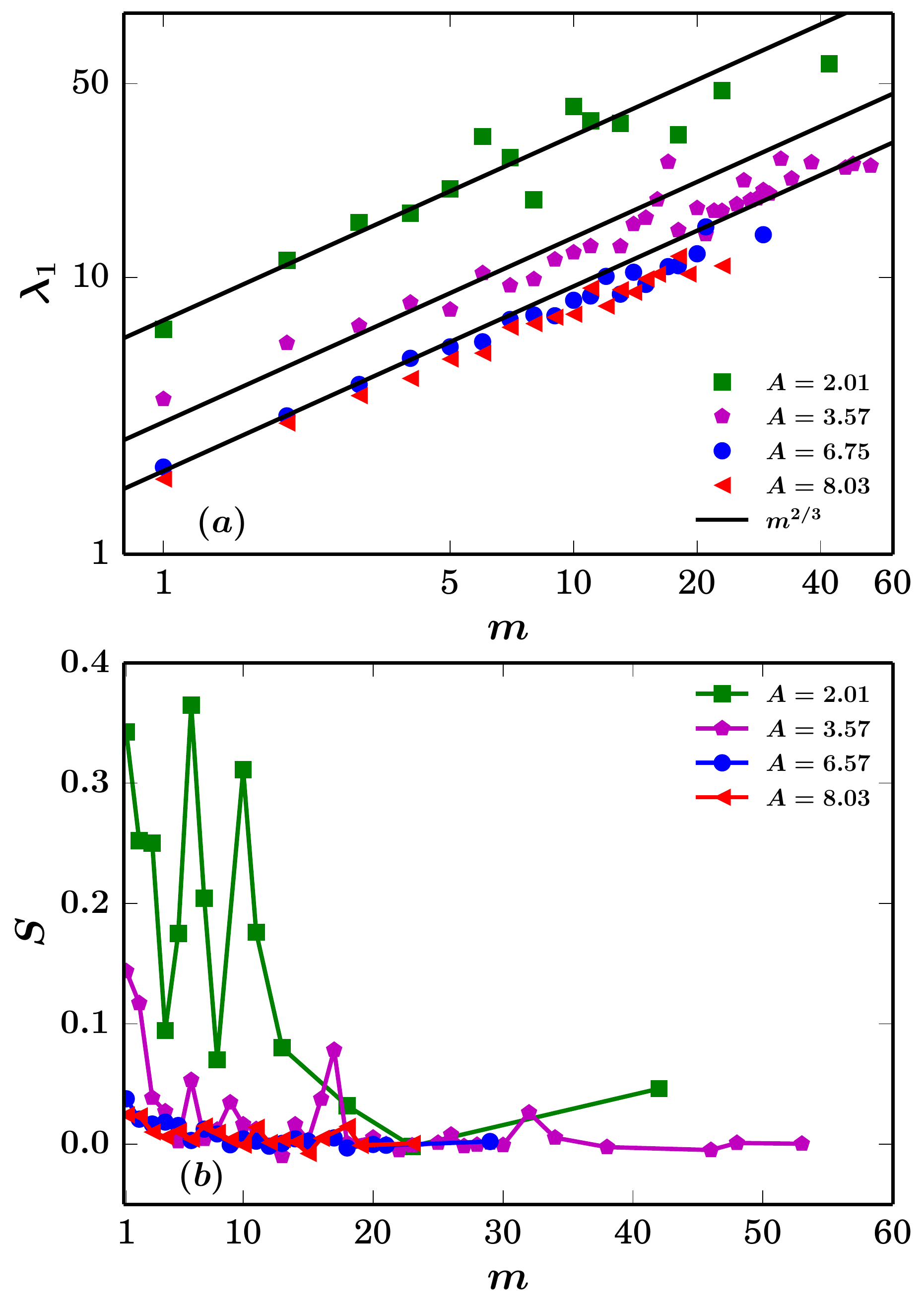} 
\caption { (a) The variation of the largest eigenvalue $\lambda_1$ of the gyration tensor with aggregate size $m$ for different charge densities $A$. (b) The variation of the prolateness $S$ with $m$ for different values of $A$.}
\label{Fig:E}
\end{figure}

In our previous work on RLPE chains~\cite{anvy16}, it was observed that the morphology of the aggregates depended on the charge density $A$ of the individual RLPE chains. For lower values of $A$, the aggregates were cylindrical with length same as that of a single RLPE chain. However, at higher values of $A$, the aggregates were more elongated, with the RLPE chains attaching to each other end-to-end. In this section, we examine the effect of backbone flexibility on the morphology of the aggregates of FPE chains. The aggregates of FPE chains are compact as may be seen in Fig.~\ref{Fig:Aggphases} and we use the gyration tensor, defined below, and the corresponding eigenvalues to characterize their shape.  
The gyration tensor is defined as
\begin{equation}
G_{\alpha\beta}=\frac{1}{N}\sum _{i=1} ^N r_{i\alpha} r_{i\beta},
\label{eq:gyration}
\end{equation}
where $r_{i\alpha}$ is the $\alpha^{\rm{th}}$ component of the position vector $\vec{r}_i$. The three eigenvalues of this tensor, 
$\lambda _i$, $i=1,2,3$ with $\lambda _1 >\lambda _2 >\lambda _3$ measure the square of the aggregate dimensions
along the three principal axes, and  can be used to define parameters that 
reflect the morphology of the aggregate. In the case of a compact morphology, the radius is proportional to $m^{1/3}$, where $m$ is the size of the aggregate and hence the eigenvalues are expected to scale as $m^{2/3}$. The variation of the largest eigenvalue $\lambda_1$ with the size of the aggregate, $m$, for different values of $A$ is shown in Fig.~\ref{Fig:E}(a) and for all values of $A$ shown, $\lambda_1$ is proportional  to $m^{2/3}$, as expected. The dependence of $\lambda_2$ and $\lambda_3$ on aggregate size $m$ is similar.

We now determine the deviation of the shape of aggregates from a perfect sphere by measuring
prolateness, a measure of the spheroidness of an object. Prolateness, in terms of the eigenvalues of the gyration tensor defined in Eq.~(\ref{eq:gyration}), is given by
\begin{equation}
S=\prod_{i=1}^3\frac{ \lambda _i -\langle\lambda\rangle}{\langle\lambda\rangle},
\label{eq:prolateness}
\end{equation}
where $\langle\lambda\rangle$ is the average value of $\lambda _i$. 
For a perfect sphere $S=0$, while for an ellipsoid, $S>0$.
The dependence of $S$ on $A$ and size of the aggregate $m$ are shown in Fig.~\ref{Fig:E}(b). It can be seen that as the size of the aggregate increases, for all values of $A$, the shape becomes more spherical and $S$ approaches 0. 

\begin{figure}
\includegraphics [width=\columnwidth]{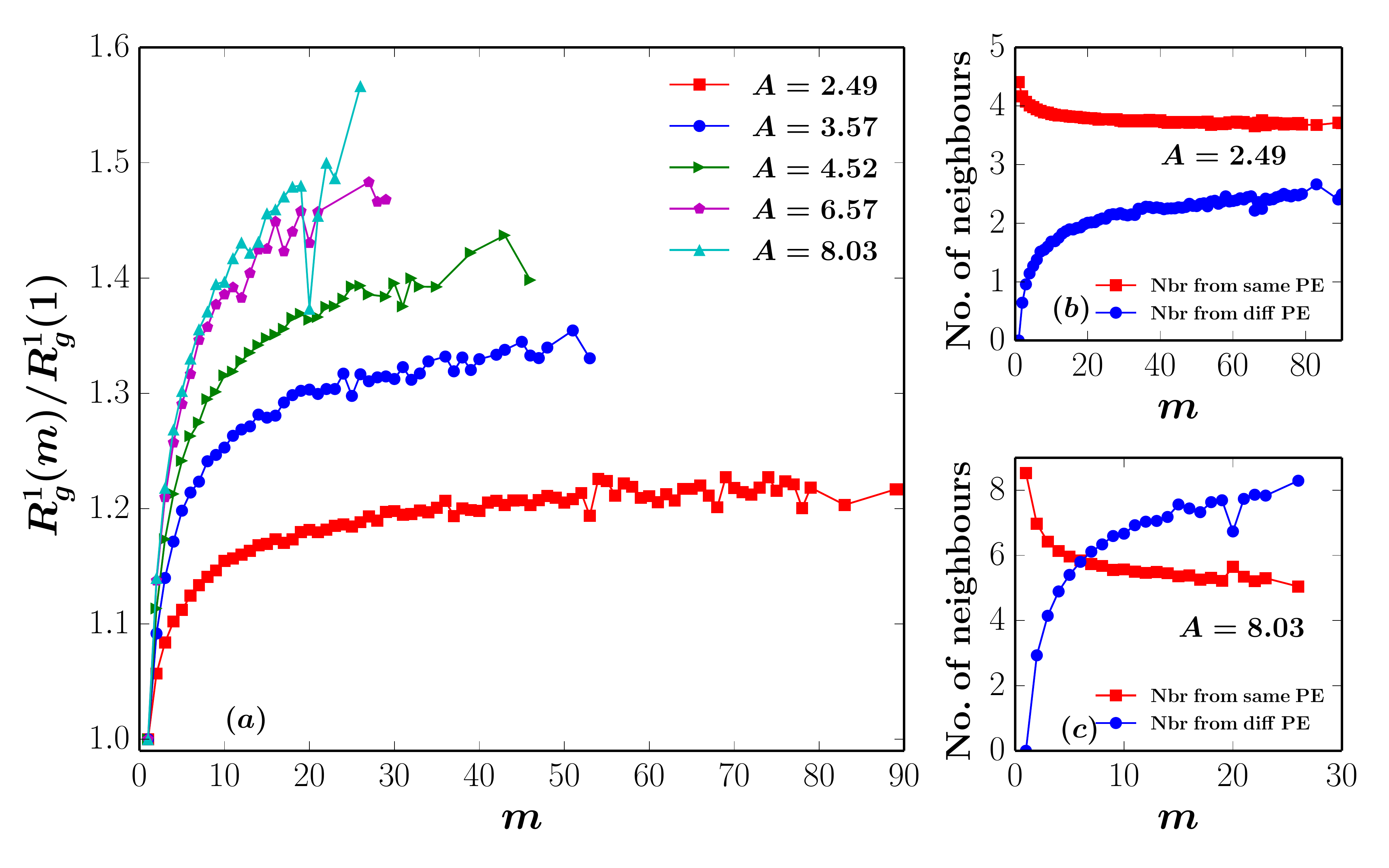}
\caption {(a)Scaled radius of gyration of a single FPE chain belonging to an aggregate of size $m$, as a function of $m$ for different $A$ for the forward simulation. The average number of non-bonded nearest neighbors of a monomer, that belong to same/different FPE chains for (a) $A=2.49$ and (b) $A=8.03$. }
\label{Fig:densitym}
\end{figure}

Next, we probe the conformations of individual FPE chains inside an aggregate in order to understand their role to act as limiting factors in the path to achieve complete phase separation at high values of $A$. Towards this goal, we calculate $R_g^1(m)$, the average radius of gyration of a single PE chain in an aggregate of size $m$ relative to its gyration radius when it is not part of any aggregate, $R_g^1(1)$, as a function of $m$ as well as $A$ and the data are shown in  Fig.~\ref{Fig:densitym}(a). The gyration radius of a FPE chain increases for all values of $A$, implying that a single FPE chain is in a more extended state in an aggregate. The relative change is larger for higher values of $A$. The implications of such extended states of single FPE chains in aggregates can be the emergence of strong entanglement of individual chains in the aggregate, which can potentially slow down the aggregation dynamics at higher values of $A$ as will be discussed in next section.  To quantify the entanglement of individual FPE chains in aggregates, we compute the average number of non-bonded nearest neighbors (within a radius of $2\sigma$) for any monomer of a FPE chain and ask whether they belong to the same FPE chain or a different one. In the case of high entanglement (and more extended conformations), we expect that any given monomer will see more neighbors belonging to other chains than its own chain. From data shown in Fig.~\ref{Fig:densitym}(b) and (c), it is clear that at low values of $A$, an individual FPE chain is more likely to be in a more compact conformation, with any monomer seeing more neighbors belonging to its own chain and even as the size of  the aggregate increases, the nature of the nearest neighbors remains the same. However, at high values of $A$, there is a crossover regime, as a function of aggregate size, where any given monomer of an individual FPE chain begins to see more neighbors from other chains than from its own polymer chain. This, when correlated with the data in Fig.~\ref{Fig:densitym}(a), shows that at high values of $A$, the individual chains are not only in more extended conformation, but are also entangled with other chains.   

\begin{figure}
\includegraphics[width=\columnwidth]{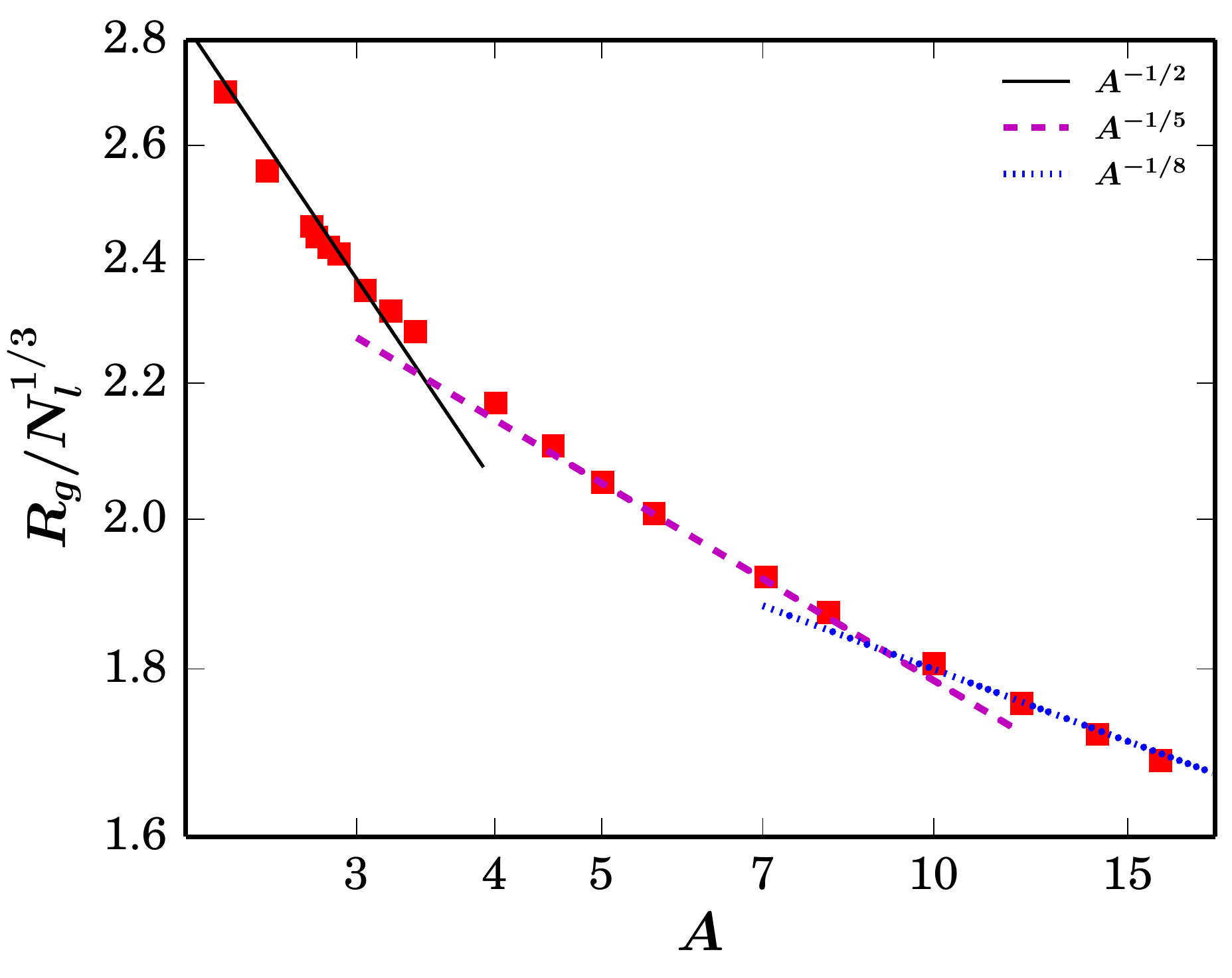} 
\caption {The scaling of the gyration radius $R_g$ of largest aggregate with charge density $A$. Each point in the plot is averaged over 2000 frames. $N_l$ is the size of largest aggregate.}
\label{Fig:RgvsA} 
\end{figure}

The entanglement of FPE chains in an aggregate suggests that an aggregate of $m$ FPE chains of length $N_m$ behaves like a single FPE chain of size $m \times N_m$. In earlier work~\cite{tom2016mechanism,tom2016regimes}, 
we showed that the collapsed state of a single FPE chain consists of multiple sub-regimes. These power-law sub-regimes are characterized by 
different scaling relations between $R_g$ and $A$, depending on the most dominant volume interaction viral term in the 
free energy expression for a collapsed FPE chain with electrostatics described by counterion fluctuation theory. 
For aggregates of FPE chain with high values of $A$ it can be envisaged that the counterions inside the aggregate may not have a preference to a particular FPE chain and may behave collectively as counterions inside a collapsed phase of a single long FPE chain. 
If this is true, we should be able to observe sub-regimes, similar to that for a single PE chain, as the value of $A$ is increased. To check this, we measured $R_g/N^{1/3}$  for the largest aggregate at a particular value of $A$, from our reverse simulations and are shown in Fig.~\ref{Fig:RgvsA}.  For small values of $A$, $R_g$ decreases as $R_g \sim A^{-1/2}$, while for large values of $A$,
$R_g$ decreases as $R_g \sim A^{-1/5}$, and for even larger values of $A$, the scaling is consistent with $R_g \sim A^{-1/8}$. These sub-regimes are identical to those obtained in simulations of
a single PE chain and as predicted by the counterion fluctuation theory for the effective attractive interactions between monomers
of the PE chain~\cite{tom2016mechanism,tom2016regimes}. This clearly shows that once, two PE chains aggregate, the aggregate may be treated as
a single PE chain of twice the original length. 

\subsection{Dynamics \label{sec:dynamics}}

We now study the pathway to full phase separation (for high values of $A$) 
starting from an initial condition of completely dispersed PE chains, using forward simulations, as described in Sec.~\ref{sec:model}. Here, we focus on three aspects of the pathway, which have been also studied for RLPE chains~\cite{henle2005equilibrium,Ha,Fazil2007,Fazil2009,Holm2007,Holm2010,savelyev,anvy16,anoop12}. 
First, the aggregation dynamics are characterized by monitoring the evolution of fraction of aggregates $n(t)$ with time. 
For RLPE chains, $n(t)$ decreases
as a power law $t^{-\theta}$, showing the absence of a characteristic size of aggregates. Using extensive MD simulations~\cite{anoop12,anvy16}, we showed that $\theta$ is independent of system parameters such as charge density $A$ of the RLPE chains, valency of counterions etc.,
and has a value $\theta\approx 0.62$, in contrast to $\theta=1$
obtained in other studies~\cite{Fazli2,Fazli1}. Second aspect is whether the aggregation dynamics can be understood by modeling  the system using classical 
Smoluchowski coagulation equation, which describes irreversible aggregation of particles that are transported  by some process, such as
diffusion or ballistic motion, and aggregate on contact~\cite{leyvraz2005,ccareview}. By modeling the RLPE chains as neutral, rotating cylinders that diffuse and aggregate on contact, we obtained~\cite{anvy16} $\theta=2/3$, in close agreement with the results from MD simulations ($\theta\approx 0.62$), strongly suggesting that  even though the dominant Coulomb interactions in this system of charged PE chains are inherently long-ranged in nature, the effective interactions that drive the aggregation process are short-ranged in nature. The third  aspect of the aggregation pathway that is of interest is understanding the mode of merging of two aggregates. Simulations suggest that, at lower values of $A$, two merging RLPE chains approach each other perpendicularly, 
followed by a sliding motion of one of the RLPE chains onto the other in a manner that has been referred to as zipper model~\cite{anoop12,anvy16,Fazli2,Fazli1,Holm2007,Holm2010,Nguyen02}.  At higher values of $A$, the mode of aggregation of two aggregates changes from a zipper model to a collinear model where the 
approaching RLPE chains are parallel to each other and join end to end with larger sliding times, resulting in elongated structures spanning the simulation box~\cite{anvy16}.  However, irrespective of the mode of aggregation, it was observed that the aggregate size increases monotonically and no fragmentation occurs in the pathway towards a bigger aggregate structure from constituent smaller aggregates.  In this section, we examine how the flexibility of the FPE chains affect the three aspects of aggregation dynamics described above.
\begin{figure}
\includegraphics [width=\columnwidth]{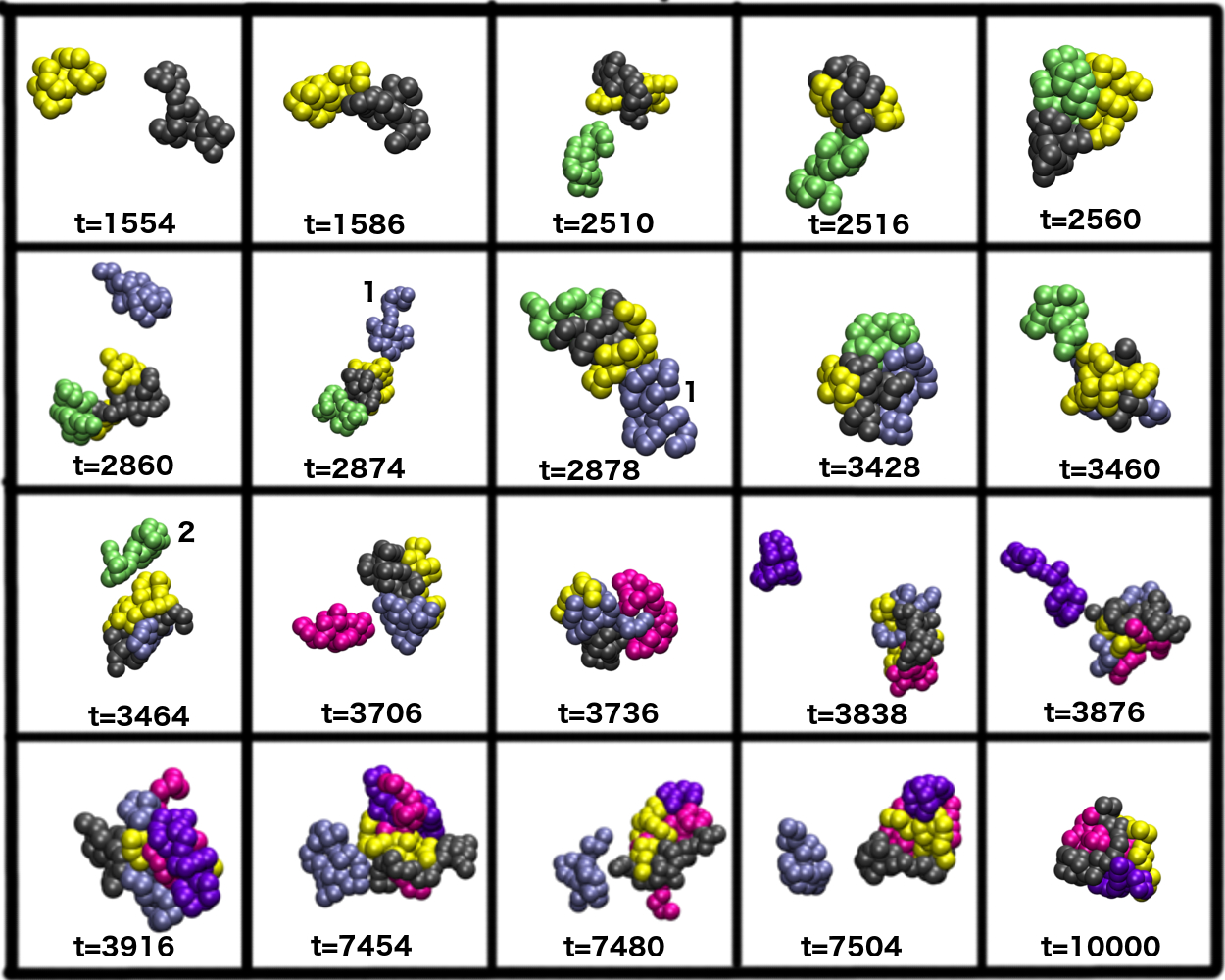} 
\caption {Snapshots of aggregation and fragmentation events over a time period for an aggregate of size $m=4$ for $A=3.57$.}
\label{Fig:cluster4}
\end{figure}

In case of FPE chains, unlike the case of RLPE chains,  we observe fragmentation events along the pathway of aggregation formation for intermediate values of $A$. Snapshots of a typical aggregation pathway of formation of an aggregate with 4 FPE chains are shown in  Fig.~\ref{Fig:cluster4} for $A=3.57$. It clearly shows intermittent fragmentation events at $t=3464$, and $7480$, where we identify 
a fragmentation event as one in which a PE chain that was a  long-existing component of an aggregate gets separated. A fragmentation
event  is usually preceded by the addition of a new FPE chain into the aggregate. For example, in the fragmentation event around $t=3464$, 
it can be noticed that the inclusion of a new FPE chain, labeled 1, results in the subsequent expulsion of an already existing FPE chain of the aggregate, labeled 2.  We, however, note that the fragmentation events are less frequent as the charge density of the FPE chains increases, possibly due to high attractive electrostatic interactions among the constituent FPE chains in an aggregate, and the higher entanglement between FPE chains as discussed in Sec.~\ref{sec:morphology}.

\begin{figure}
\includegraphics [width=\columnwidth]{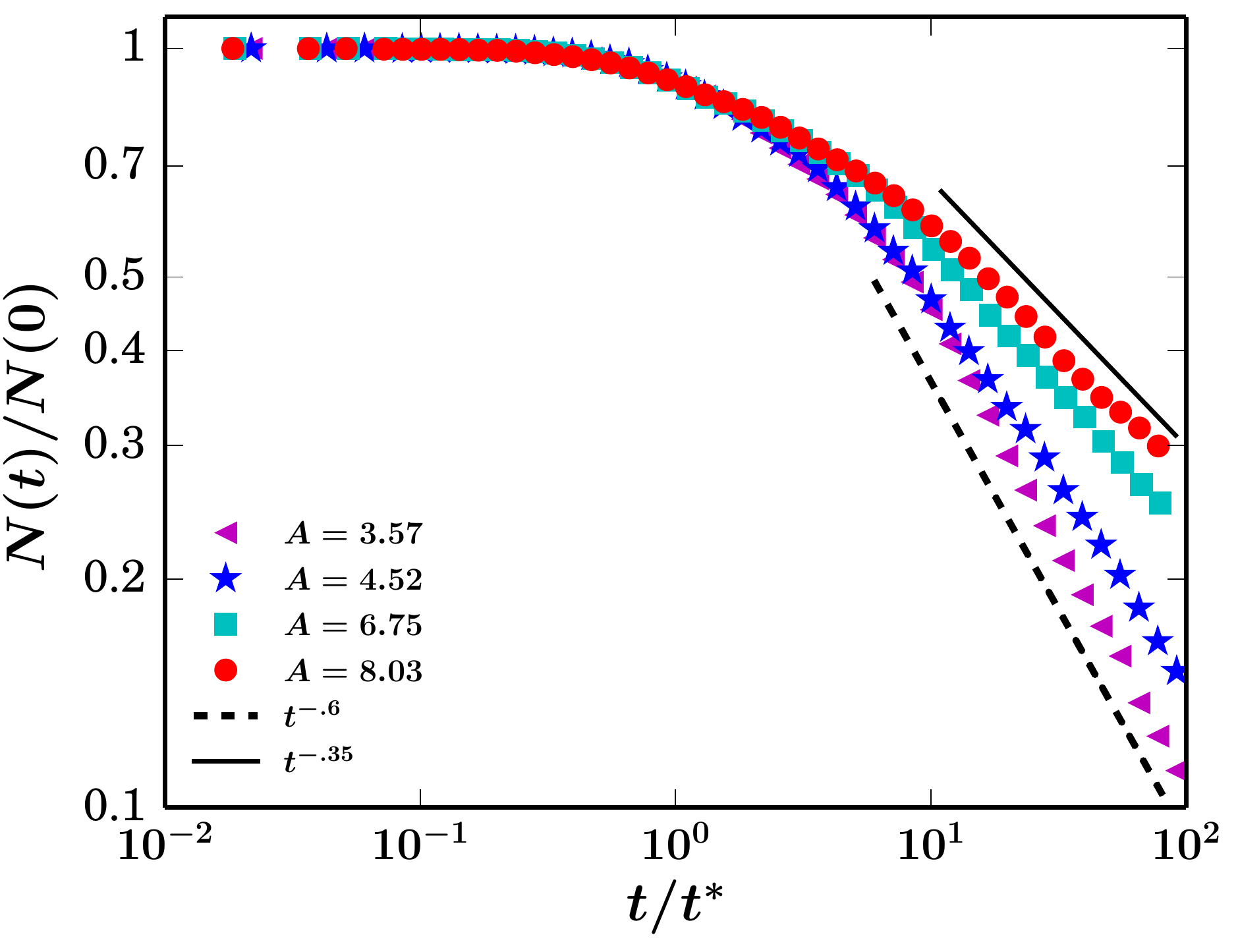}
\caption {The variation of fraction of aggregates $n(t)$ with scaled time $t/t^*$ for different values of $A$, where $t^*$ is the time at which $n(t)=.9\times n(0)$.The straight lines are power laws and guides to the eye.}
\label{Fig:fractionn}
\end{figure}

The aggregation dynamics is characterized by monitoring the temporal evolution of fraction of aggregates, $n(t)$, which decreases on aggregation and increases with fragmentation and attains a value of $N^{-1}=0.01$, in the fully phase segregated phase or remains close to $1$ in the non-aggregated phase. The variation of $n(t)$ with scaled time $t/t^*$ is shown  in Fig.~\ref{Fig:fractionn}, where $t^*$ is set to be the time when $n(t)=0.9 n(0)$. It can be seen that $n(t)$ decreases with $t/t^*$ as a power law, albeit with varying values of the exponent $\theta$ which decrease with the charge density of the FPE chain, in stark contrast to the single exponent ($\theta\approx0.62$) that we obtained in the aggregation dynamics of RLPE chains. The exponent $\theta$ varies from $0.6$ to $0.35$ for the range of $A$ values considered here.   

We now ask whether the aggregation dynamics of the FPE chains can be modeled by the Smoluchowski equation, 
which was successfully applied to the case of aggregation of RLPE chains.
The Smoluchowski equation for irreversible aggregation (for reviews, see~\cite{leyvraz2005,ccareview}) is 
\begin{align}
\frac{dN(m,t)}{dt}&= \frac{1}{2}\sum_{m_1=1}^{m-1} K(m_1,m-m_1) N(m_1)
N(m-m_1) \nonumber \\
& -\sum_{m_1=1}^\infty K(m,m_1) N(m) N(m_1),
\label{eqn1}
\end{align}
where $N(m,t)$ is the number of aggregates of size $m$ at time $t$,
and $K(m_1,m_2)$ is the rate at which two aggregates $m_1$ and $m_2$
collide. The first term in Eq.~(\ref{eqn1}) describes the 
aggregation of particles to form an aggregate of size $m$, while the
second term  describes the loss of an aggregate of size $m$ due to
collision with another aggregate. 
If the collision kernel $K(m_1,m_2)$ is a homogeneous function of its arguments
with homogeneity exponent $\lambda$, i.e., $K(hm_1,h m_2) = h^\lambda
K(m_1,m_2)$, then the number of aggregates $N(t) = \sum_m N(m,t)$,
decreases in time as a power law $N(t) \sim t^{-\theta}$, where
\be
\theta=\frac{1}{1-\lambda}, \quad \lambda<1.
\label{eq:beta}
\ee
For diffusing spheres in three
dimensions, the coagulation kernel is known to be (for example, see~\cite{krapivsky2010kinetic,leyvraz2005})
\begin{equation}
K(m_1,m_2) \propto [D(m_1)+D(m_2)][R(m_1)+R(m_2)],
\label{eq}
\end{equation}
where $D(m)$ and $R(m)$ are the diffusion constant and effective radius of an
aggregate of $m$ PE chains.
In the absence of a solvent, the diffusion constant is inversely
proportional to its size:
$D(m) \propto m^{-1}$. As can be seen from Fig.~\ref{Fig:Aggphases} and Fig.~\ref{Fig:E}, the shape of aggregates, in the case of FPE chains, is spherical and 
 hence $R(m) \propto m^{1/3}$, giving $\lambda=-2/3$. Substituting this value of $\lambda$ for spherical aggregates in Eq.~(\ref{eq:beta}),
we obtain $\theta=3/5$. This is in contrast to the value of $\lambda=-1/2$ and $\theta=2/3$ that we obtained in the case of RLPE chains~\cite{anvy16}, using the kernel for rotating cylindrical aggregates with $R(m) \propto m^{1/2}$.
\begin{figure}
\includegraphics [width=\columnwidth]{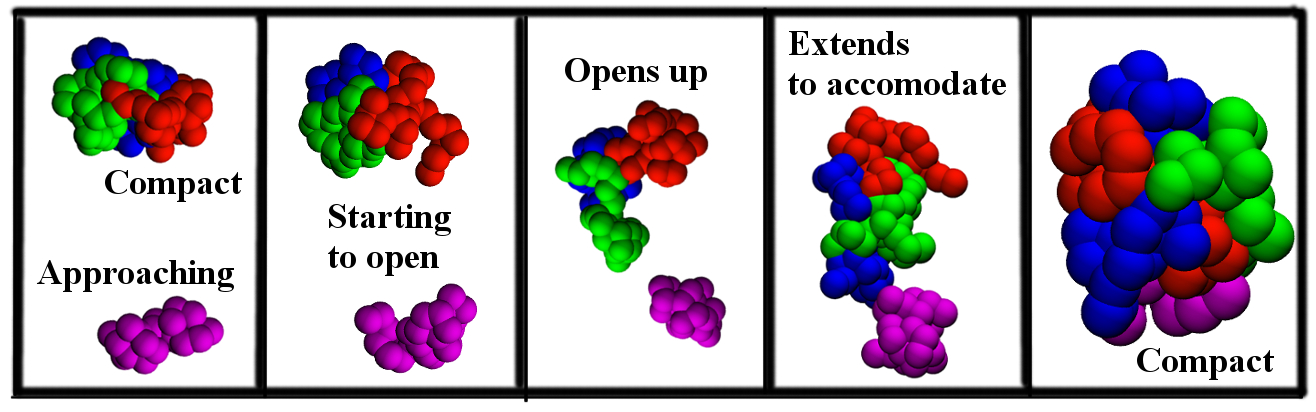} 
\caption {Snapshots of aggregation showing the opening up of an existing aggregate to incorporate another PE chain.}
\label{Fig:clusterformation}
\end{figure}

The result obtained from Smoluchowski theory ($\theta=3/5$) is close to the numerical value obtained
for smaller values of $A$ (see lower guide line in Fig.~\ref{Fig:fractionn}). However, the exponent for larger
values of $A$ is dependent on $A$ and is significantly smaller  $\theta=3/5$ (see upper guide line in Fig.~\ref{Fig:fractionn})
showing   a lower aggregation rate with larger $A$. This discrepancy could possibly be due to 
the existence of kinetic barriers, due to the entanglement of the extended FPE chains in aggregates, that have to be crossed when two aggregates merge. In the case of RLPE chains, the typical aggregate conformation is linear and a new RLPE chain gets attached 
to the existing aggregate without modifying the structure of the previous aggregate. This is not the case
for FPE chains, as may be seen from the 
snapshots of a typical aggregation event  shown in Fig.~\ref{Fig:clusterformation}.  In the case of FPE chains,  owing to their inherent highly 
flexible backbones, the aggregates have a more entangled structure and incorporation of a new FPE chain into such an aggregate requires significant global conformational changes, as can be seen in Fig.~\ref{Fig:clusterformation}.
This process can introduce additional time scales into the aggregation dynamics, altering the power law exponent from the result obtained via Smoluchowski theory suggesting that the flexibility of the PE chain backbone can significantly
alter the aggregation dynamics as compared to RLPE chains. It can also be envisaged that these additional time scales become increasingly more significant at higher values of $A$, where the strong electrostatic interactions are 
the dominant forces holding the aggregate together leading to larger deviations from the classical Smoluchowski theory.

\section{Summary and discussion  \label{sec:summary}}

In this paper we determined, using extensive MD simulations, the effects of backbone flexibility of FPE chains on their aggregation dynamics as well 
as the equilibrium phases in the presence of trivalent counterions, and contrasted them with the known results for RLPE chains with rigid backbones. We showed
the existence of three possible phases. At low charge densities, the FPE chains do not aggregate and remain uniformly distributed throughout the available
volume. At intermediate charge densities, the system exists in a finite bundle phase, where a  fraction of the FPE chains form an aggregate while 
the remaining FPE chains are in the form of multiple small aggregates. At large charge densities, the equilibrium configuration is one of phase
separation where all the FPE chains come together as a single aggregate. Surprisingly the critical charge density for the transition
from finite bundles to fully phase separated phase  would appear to be independent, within numerical error, of the number density of the system. We characterized the morphology of the aggregates in detail. Due to
the flexibility of the PE chain, the aggregates are compact and their shapes become more spherical with increasing charge density as well as
aggregate size. An individual FPE chain within an aggregate becomes more extended with increasing aggregate size, and we show that its non-bonded nearest neighbors are 
increasingly from other FPE chains, implying that the FPE chains within an aggregate are strongly entangled. 
Finally, we characterized the dynamics of aggregation in the phase separated phase by measuring the power law exponent describing the 
decreasing in the number of aggregates. Surprisingly, we find this exponent to vary with  charge density.

The equilibrium phases of aggregates of PE chains has been a long standing issue,
with most of the studies  focusing on RLPE chains~\cite{Fazil2007,Fazil2009,Holm2007,Holm2010,savelyev}.
Conflicting results from both experimental and theoretical studies suggest that the final equilibrium configuration of an aggregate phase, 
in the presence of multivalent counterions, could be either a fully phase separated state or coexisting finite aggregates~\cite{Fazil2007,Fazil2009,Holm2007,Holm2010,savelyev}. 
Theoretical studies have also shown that the size of the multivalent counterion, frustration of the  interactions between the rods due to growth of the 
aggregate, as well as kinetic barriers  play a role in determining whether finite-sized bundles or complete phase separation are the 
equilibrium states~\cite{henle2005equilibrium,Ha}. However, by performing large-scale simulations
on systems of 100 RLPE chains, unlike earlier simulations with fewer PE chains as well as shorter simulation time scales~\cite{Fazil2007,Fazil2009,Holm2007,Holm2010}, we showed that  the equilibrium state is either a 
fully separated state or one of no aggregation, and could find no evidence
of a finite bundle phase~\cite{anvy16}. In contrast, for FPE chains, we find that for a large regime of charge density for each value of number density, there exists a finite bundle phase where
 large aggregates coexist with multiple small aggregates, emphasizing the crucial role played by the flexibility of the polymers. In the 
fully separated phase, there is always a finite probability of FPE chains fragmenting and forming smaller aggregates. However, we expect that
this fraction would be thermodynamically negligible. To show this more rigorously, we would need to systematically simulate larger systems, which is
computationally expensive and is beyond the scope of the present study.

The aggregation dynamics of the FPE chains is qualitatively and quantitatively different from that of  RLPE chains. Qualitatively, we find that in the aggregation
of FPE chains, we observe fragmentation events where an existing aggregate breaks up into smaller ones. This occurs particularly for charge
densities corresponding to finite bundles phase. Such fragmentation events were not observed in the aggregation of RLPE chains~\cite{anvy16}. The aggregation
dynamics were quantified through the temporal variation of the fraction of aggregate $n(t)$ in forward simulations: $n(t) \sim t^{-\theta}$. We find 
that for FPE chains, $\theta$ decreases with increasing charge density and varies from from $0.6$ and $0.35$ for the range of charge densities considered.
This is in complete contrast to the charge independent value of $\theta \approx 0.62$  obtained for  RLPE chains through MD simulations~\cite{anvy16}.  
We rationalize the additional time scales that emerge by carefully monitoring a single aggregation event in which a new FPE chain merges with the 
existing aggregate. To incorporate a new FPE chain into an aggregate,  a global reorganization of the existing aggregate is required in the form of adopting an 'open' configuration as the new FPE chain approaches.  Once the FPE chain is assimilated into the existing aggregate, another global conformational rearrangement occurs before the new aggregate adopts an almost spherical shape and 'closed' conformation. This opening and closing of the structures can have two possible consequences. 
If the effective attractive interactions holding the FPE chains in the aggregate are not strong enough, fragmentation events can occur slowing down the aggregation dynamics process. Additionally, at higher charge densities, the opening of the aggregate structures may be difficult due to high attractive energetic barriers introducing additional time scales, issues that do not impact the aggregation of the RLPE chains. The slowing down of aggregation dynamics at higher values of $A$, for FPE chains, can also be understood in terms of individual conformations of FPE chains in aggregates. We showed that as the value of $A$ increases, the individual FPE chains in aggregates are more extended, compared to the case of a single FPE chain, and are also more entangled with other chains resulting in appearance of kinetic barriers in the pathway of forming bigger aggregates. The stronger entanglement  at larger values of $A$ also explains why fragmentation events occur only at lower values of $A$.  

Understanding the effective interactions driving both the collapse of a single PE chain and aggregation of similarly charged PE chains is a long-standing problem of fundamental interest. In particular, the emergence of short-ranged attractive interactions in a system of similarly charged entities with inherently long-ranged Coulomb interactions is an aspect that is not well-understood. Several theories~\cite{brilliantov98,pincus98,Kardar1999,delaCruz2000} have been proposed to explain the nature of the effective interactions based on minimization of free energy of a PE chain. In our earlier work~\cite{tom2016mechanism,tom2016regimes}, using both extensive MD simulations and analytical calculations, we showed conclusively that the counterion fluctuation theory and its modifications best explains the observed emergence of multiple regimes in the collapse phase of a single PE chain. The data Fig.~\ref{Fig:RgvsA} shows the emergence of similar regimes, characterized by different scaling exponents in the relation between radius of gyration and the effective Bjerrum length of the PE chain ($R_g\sim A^{-\alpha}$) as predicted by counterion fluctuation theory for a single PE chain collapse.  This is further evidence for the counterion fluctuation theory being the correct description for the effective attractive interactions in charged PE systems. The results also suggest that at high values of $A$, the multiple FPE chains in an aggregate are strongly entangled and behave effectively like a single long PE chain. 

Considering our observation that the aggregates in case of FPE chains are nearly spherical, we model the aggregation of FPE chains using Smoluchowski equation, which for neutral spheres that diffuse and coagulate on contact, gives $\theta=3/5$ for aggregation dynamics. In the case of FPE chains we obtained a charge dependent value of $\theta$ which varied from $0.6$ and $0.35$. To obtain such a charge dependent $\theta$, it would be necessary to modify the collision kernel to include effects
beyond geometry, such as the time scale associated with opening up an aggregate consisting of entangled FPE chains. Quantifying these time scales in a systematic manner is a promising area for future study. We note that modeling the aggregation dynamics of RLPE chains using Smoluchowski coagulation equation gives $\theta=2/3$ very close to the numerical obtained value of $0.62$, and indicates the lack of rearrangement required to incorporate a new RLPE chain into an existing aggregate. It may still be possible to understand the different phases in terms of competition between aggregation and fragmentation of neutral aggregates.  
In simple lattice models of aggregation and fragmentation, an interesting phase transition occurs when fragmentation is limited to a finite number of
units fragmenting from an existing aggregate to a neighbor~\cite{MKB1,MKB2,krapivsky1996transitional,rajesh2001exact,rajesh2002effect,rajesh2002aggregate}.
As fragmentation rate is decreased, the system undergoes a non-equilibrium phase transition from a phase 
characterized by an exponential distribution of aggregate sizes (akin to no aggregation) to a
phase characterized by the presence of a condensate containing a finite fraction of the polymers, and the remaining polymers  being in smaller
aggregates distributed as a power law. In the limit of zero fragmentation, phase separation happens  in the form of a single condensate.
The phases seen in this paper have a one to one mapping with the phases seen in such an aggregation-fragmentation model. However, the mapping
is at a qualitative level and understanding the role of charge density in determining the fragmentation rate in systems like charged polymers is an interesting open problem.

\begin{acknowledgments}
The simulations were carried out on the 
supercomputing
machines Annapurna, Nandadevi and Satpura at The Institute of Mathematical Sciences.
\end{acknowledgments}

%\bibliography{FPE-agg2} 
%merlin.mbs aipnum4-1.bst 2010-07-25 4.21a (PWD, AO, DPC) hacked
%Control: key (0)
%Control: author (8) initials jnrlst
%Control: editor formatted (1) identically to author
%Control: production of article title (-1) disabled
%Control: page (0) single
%Control: year (1) truncated
%Control: production of eprint (0) enabled
%

\end{document}